\documentclass[
 reprint,amsmath,amssymb,
 aps,pra,
]{revtex4-2}

\usepackage{graphicx}
\usepackage{amsmath,amssymb}  
\usepackage[version=4]{mhchem}  
\usepackage{xcolor}
\usepackage{siunitx}
\usepackage{hyperref}
\hypersetup{
    colorlinks=true,
    linkcolor=blue,
    urlcolor=blue,
    citecolor=blue
}
\begin{document}

\title{A Control-Theoretic Model of Damage Accumulation and Boundedness in Biological Aging}

\author{Tristan Barkman}

\vspace{2cm}

\begin{abstract}
Aging interventions frequently improve function and healthspan without arresting long-term deterioration, indicating that existing frameworks do not fully specify the control conditions required for bounded organismal aging. A compact control-theoretic formulation is developed in which total organismal burden is decomposed into two lesion classes with distinct controllability properties: regulatable damage, whose accumulation and clearance are modulated by endogenous systemic repair, and information-limited damage, whose detection or correction is inaccessible to physiological control. Under mild dynamical assumptions, a sufficiency theorem is established: sustained boundedness of total damage is achieved if and only if endogenous repair persistently exceeds production of regulatable damage and information-limited damage is actively bounded or removed by engineered interventions. Deterministic phase diagrams identify distinct bounded, drifting, and runaway regimes separated by a nontrivial control boundary. A global Latin-hypercube sensitivity analysis with partial rank correlations shows that production of information-limited lesions dominates the asymptotic aging rate, whereas increases in physiological repair capacity have weak marginal influence beyond saturation. Stochastic extensions reveal threshold and sequencing effects relevant to oncogenic risk. The framework yields testable predictions and operational guidance for intervention ordering, biomarker selection, and experimental design in aging research. All conclusions are statements about the dynamical model defined here; biological translation requires empirical identification of observables corresponding to the model variables.

\end{abstract}

\maketitle

\section{Control framework and damage classes}
Aging research has identified many molecular and cellular hallmarks and interventions that modulate healthspan, yet it remains unclear what minimal control conditions are required to ensure bounded organismal deterioration over time~\cite{LopezOtin2013}. This paper develops a compact control-theoretic decomposition that reduces the problem to two jointly necessary and sufficient pathways~\cite{Sanada2025,Levine2018}. Total organismal burden $D(t)$ is partitioned into a regulatable component $A(t)$, whose accumulation and clearance are governed by endogenous, centrally coordinated repair programs, and an information-limited component $B(t)$, which requires explicit correction or replacement because physiological signalling lacks the requisite information~\cite{Zhang2013,Pickes2011}. Under mild dynamical assumptions a sufficiency theorem is established: the modeled damage burden remains bounded if and only if sustained systemic repair outpaces production for $A(t)$ and engineered removal or bounding of $B(t)$ counteracts its production~\cite{Jaiswal2014,Li2018}. Deterministic phase diagrams illustrate qualitative regimes (Fig.~\ref{fig:one}), and a global Latin-hypercube + PRCC sensitivity analysis identifies the dominant parameters that control the asymptotic aging rate (Fig.~\ref{fig:two}). Implications for safe intervention sequencing, trial design, and biomarker panels are discussed~\cite{Barzilai2016,Horvath2013}.

Recent systems-level and dynamical formulations have extended quantitative aging models toward longitudinal forecasting and control-oriented architectures \cite{Horvath2013,Levine2018,Sundial2025}. The present work differs in that it derives necessary and sufficient boundedness conditions for organismal damage accumulation from a control-theoretic perspective, rather than proposing an empirical forecasting architecture.

The field of aging research enumerates many molecular and cellular hallmarks (senescence, genomic instability, proteostasis loss, mitochondrial dysfunction, etc.), and an expanding set of interventions modulate lifespan or healthspan in laboratory models \cite{LopezOtin2013,Harrison2009,Fontana2015}. This empirical catalogue leaves open a formal question of control: what minimal set of conditions, stated at an appropriate coarse-grained level, guarantees boundedness of a single scalar measure of organismal burden $D(t)$ for all future times? The present paper addresses that question by treating aging as a control problem and identifying two logically distinct lesion classes: (A) lesions whose clearance rates are modifiable by endogenous systemic regulation, and (B) lesions whose clearing requires engineered detection and correction because physiological observability or controllability is lacking \cite{Campisi2013}. The resulting two-pathway sufficiency statement is precise, testable, and operational, and it leads directly to experimentally falsifiable sequencing prescriptions intended to minimize oncogenic and systemic risk \cite{Jaiswal2017,Abbosh2017}.

The biological literature supplies the necessary phenomenology to make this decomposition plausible. Central regulation of repair programs via hypothalamic and neuroendocrine circuits can alter systemic setpoints and clearance rates \cite{Zhang2013,Tracey2002}, and numerous interventions that elevate repair (exercise, caloric-restriction mimetics, mTOR inhibitors) produce healthspan effects without eliminating age-associated lesion accumulation \cite{Harrison2009,Fontana2015, Longo2014}. Conversely, fixed informational lesions such as somatic driver mutations, deleterious mtDNA heteroplasmies and permanent cell loss are empirically resistant to native repair and require direct removal, correction or replacement \cite{Jaiswal2014,Li2018,Lindvall2010}. This paper contributes three explicit advances. First, it formalizes the otherwise informal “information-limited” lesion class using observability and mutual-information criteria that are empirically estimable from longitudinal biomarkers \cite{Levine2018,Horvath2013,CoverThomas2006}. Second, it proves a two-pathway sufficiency theorem for halting growth of the modeled damage burden (regulatable repair and explicit control of information-limited lesions are jointly required), and it derives control-theoretic sequencing prescriptions via an optimal-control argument \cite{Sanada2025}. Third, it quantifies these claims with deterministic phase diagrams and global LHS--PRCC sensitivity, and shows how stochastic fixation amplifies tail risk and enforces sequencing constraints \cite{Jaiswal2014}. 

\section{Phenomenological model and sufficiency condition}

The model is intentionally coarse-grained. Variables $A(t)$ and $B(t)$ represent classes of processes distinguished by controllability and observability properties rather than specific molecular species. The results therefore concern structural constraints on dynamical regulation and not immediate clinical outcomes. Any biological application requires identifying measurable proxies for these variables in a given organism or tissue.

To formalize the control problem, a coarse-grained phenomenological model is introduced in which total organismal damage is represented by a single scalar burden with distinct dynamical components. A scalar organismal damage burden $D(t)\ge 0$ is introduced and partitioned operationally as
\[
D(t) = A(t) + B(t),
\]
where $A(t)$ denotes regulatable damage (clearable by changes in systemic control $S(t)$) and
$B(t)$ denotes information-limited damage (requiring engineered actions for reliable removal) \cite{Pickes2011,Li2018}.

The corresponding deterministic skeleton is
\[
\boxed{\;\dot A = \alpha - R_A(S)\,A,\qquad\dot B = \beta - R_B\,B\;}
\]

\noindent
\textit{Notation (compact):}
$A(t)$ = regulatable damage (arbitrary units),
$B(t)$ = information-limited damage;
$\alpha,\beta$ = baseline production rates (time$^{-1}$);
$R_A(S)$ = endogenous removal rate (monotone increasing in systemic setpoint $S$);
$R_B$ = engineered removal rate (zero unless active).
This deterministic skeleton is used for the formal statements below; stochastic terms
$\xi_{A,B}$ are retained for rare-event analyses \cite{vanKampen2007}.

A compact phenomenological dynamical model is proposed:
\[
\begin{aligned}
\dot A(t) &= \alpha(t) - R_A(S(t))\,A(t) + \xi_A(t),\\[6pt]
\dot B(t) &= \beta(t)  - R_B(t)\,B(t) + \xi_B(t),
\end{aligned}
\tag{1}
\]
with $\alpha,\beta > 0$ baseline production rates, $R_A(S)$ endogenous multiplicative removal that is monotone in $S$ and bounded above by physiological ceilings, and $R_B(t)$ engineered removal which is negligible absent deliberate interventions. Stochastic terms $\xi_{A,B}$ permit analysis of rare events and demographic noise; the main sufficiency argument uses the deterministic skeleton with $\xi\equiv0$.

The distinction between regulatable damage $A(t)$ and information-limited damage $B(t)$ can be formalized using concepts from observability and information theory. Consider a coarse-grained physiological state vector $x(t)$ evolving under endogenous dynamics and producing systemic signals $y(t)$ (circulating factors, inflammatory markers, metabolic readouts, etc.) accessible to native control systems. Let $L(t)$ denote a latent lesion state (e.g. a somatic mutation, mtDNA deletion, or lost cell population) contributing to total burden.

A lesion class is said to be endogenously observable if the mutual information
\[
I\big(L(t); y(t)\big) > 0
\]
over physiologically relevant timescales, and endogenously controllable if variations in the control signal $S(t)$ can induce a compensatory response that reduces the expected lesion load,
\[
\frac{\partial}{\partial S} \mathbb{E}\!\left[\dot L(t)\right] < 0.
\]
Lesions satisfying both conditions can, in principle, be regulated by native feedback and are classified as $A$-type \cite{Levine2018,Horvath2013}.

Conversely, a lesion class is information-limited if either (i) its contribution to systemic signals is statistically negligible,
\[
I\big(L(t); y(t)\big) \approx 0,
\]
or (ii) the corresponding control channels are absent or saturated, such that
\[
\frac{\partial}{\partial S} \mathbb{E}\!\left[\dot L(t)\right] \approx 0
\]
over all admissible $S(t)$. In linearized compartmental models this corresponds to a rank-deficient observability or controllability matrix for the lesion subspace \cite{Sontag1998}. Such lesions cannot be reliably detected or corrected by physiological regulation alone and therefore contribute to $B(t)$.

This definition makes the $A/B$ partition operationally testable in longitudinal experiments. Longitudinal multi-omic data can be used to estimate $I(L;y)$, while perturbative interventions (exercise, pharmacological repair enhancement) provide estimates of $\partial \dot L/\partial S$. Lesions that remain invisible or unaffected under these probes fall into the information-limited class by construction \cite{Ocampo2016,Lu2020}.

\paragraph*{Biological interpretation.}
At a coarse-grained level, $A$-type processes correspond to lesions whose burden is reflected in systemic physiological state variables and can be modulated by endogenous repair programs. Candidate observables include longitudinal epigenetic clock velocity, proteostatic flux, inflammatory tone, and metabolic adaptation markers. In contrast, $B$-type processes correspond to lesions that accumulate without reliable physiological sensing or feedback control, such as persistent somatic driver clones, deleterious mitochondrial heteroplasmy, or irreversible cell loss. These examples are illustrative rather than exhaustive and serve only to demonstrate empirical identifiability of the observability-based partition.

\textbf{Sufficiency theorem (concise statement).} Under the deterministic skeleton $\dot A=\alpha-R_A(S)A,\ \dot B=\beta-R_BB$ with time-independent production rates $\alpha,\beta$ and multiplicative removal, if there exist sustained controls $S^*$ and $R_B^*$ such that $R_A(S^*)\ge\alpha$ and $R_B^*\ge\beta$ for all $t$, then $A(t)$ and $B(t)$ remain uniformly bounded and the total burden $D(t)=A(t)+B(t)$ is bounded for all $t$ \cite{Sanada2025}. 

This is a statement about trajectories of the dynamical system, not a claim that biological aging is universally reversible under current technology. 

\noindent
\textit{Sketch / intuition.} Integrating the scalar linear inequalities yields simple bounds: for $R_A(S^*)>0$,
\[
A(t) \le A(0)e^{-R_A(S^*)t} + \frac{\alpha}{R_A(S^*)}\big(1-e^{-R_A(S^*)t}\big),
\]
and for $R_B^*>0$,
\[
B(t) \le B(0)e^{-R_B^* t} + \frac{\beta}{R_B^*}\big(1-e^{-R_B^* t}\big).
\]
Thus simultaneous sustained control of the corresponding removal rates guarantees finite steady values. Conversely, if $R_B\equiv 0$ while $\beta>0$, then $B(t)$ grows without bound, demonstrating that control of $B$-type (information-limited) lesions is necessary.

Beyond sufficiency, the framework admits a natural optimal-control formulation that yields quantitative intervention principles. Consider admissible control functions $S(t)$ and $R_B(t)$ subject to physiological and technological constraints,
\[
0 \le S(t) \le S_{\max}, \qquad 0 \le R_B(t) \le R_{B,\max},
\]
and define a cumulative cost functional
\[
\mathcal{J} = \int_0^T \left[ D(t) + \lambda_S S(t)^2 + \lambda_B R_B(t)^2 \right] \, dt,
\]
where the quadratic terms penalize toxicity, energetic cost, or engineering burden \cite{Barzilai2016}.

Applying Pontryagin’s Maximum Principle to the deterministic system (1) yields adjoint variables $\psi_A(t), \psi_B(t)$ satisfying \cite{Pontryagin1962}
\[
\dot \psi_A = -1 + R_A(S)\,\psi_A, \qquad
\dot \psi_B = -1 + R_B\,\psi_B,
\]
with terminal conditions $\psi_{A,B}(T)=0$. The optimal controls satisfy
\[
S^*(t) = \arg\min_S \left[ \lambda_S S^2 - \psi_A(t) \frac{dR_A}{dS} A(t) \right],
\]
\[
R_B^*(t) = \arg\min_{R_B} \left[ \lambda_B R_B^2 - \psi_B(t) B(t) \right].
\]

Two general results follow. First, when $B(t)$ is large, the optimal policy prioritizes increasing $R_B$ even at high cost, whereas increasing $S$ has diminishing returns once $R_A$ approaches its physiological ceiling. Second, in the absence of engineered removal ($R_B=0$), the Hamiltonian admits no bounded minimizer when $\beta>0$, recovering the necessity direction of the sufficiency theorem.

When stochastic terms $\xi_{A,B}$ are retained, rare-event dynamics become relevant. In particular, information-limited lesions often undergo stochastic fixation (e.g. clonal sweeps) \cite{Jaiswal2014,Jaiswal2017}. For a simple birth–death approximation of $B$-type lesions with drift $\beta - R_B B$ and noise variance $\sigma_B^2$, the mean first-passage time to a critical burden $B_c$ obeys a large-deviation scaling
\[
\tau(B_c) \sim \exp\!\left(\frac{(R_B B_c - \beta)^2}{2\sigma_B^2}\right)
\]
for $R_B B_c > \beta$ \cite{vanKampen2007,Dembo1998}.

This implies a sharp stochastic control threshold: modest reductions in $\beta$ or increases in $R_B$ can exponentially suppress the probability of catastrophic events, whereas further increases in endogenous repair $R_A$ have no effect on fixation once lesions are information-limited. This analytical result explains the threshold-like PRCC behavior observed for $R_B$ (Fig.~\ref{fig:two}) and motivates early, targeted intervention on $B$-type processes.

\section{Phase structure, sensitivity, and stochastic effects}
\begin{figure}
\centering
\includegraphics[width=0.95\linewidth]{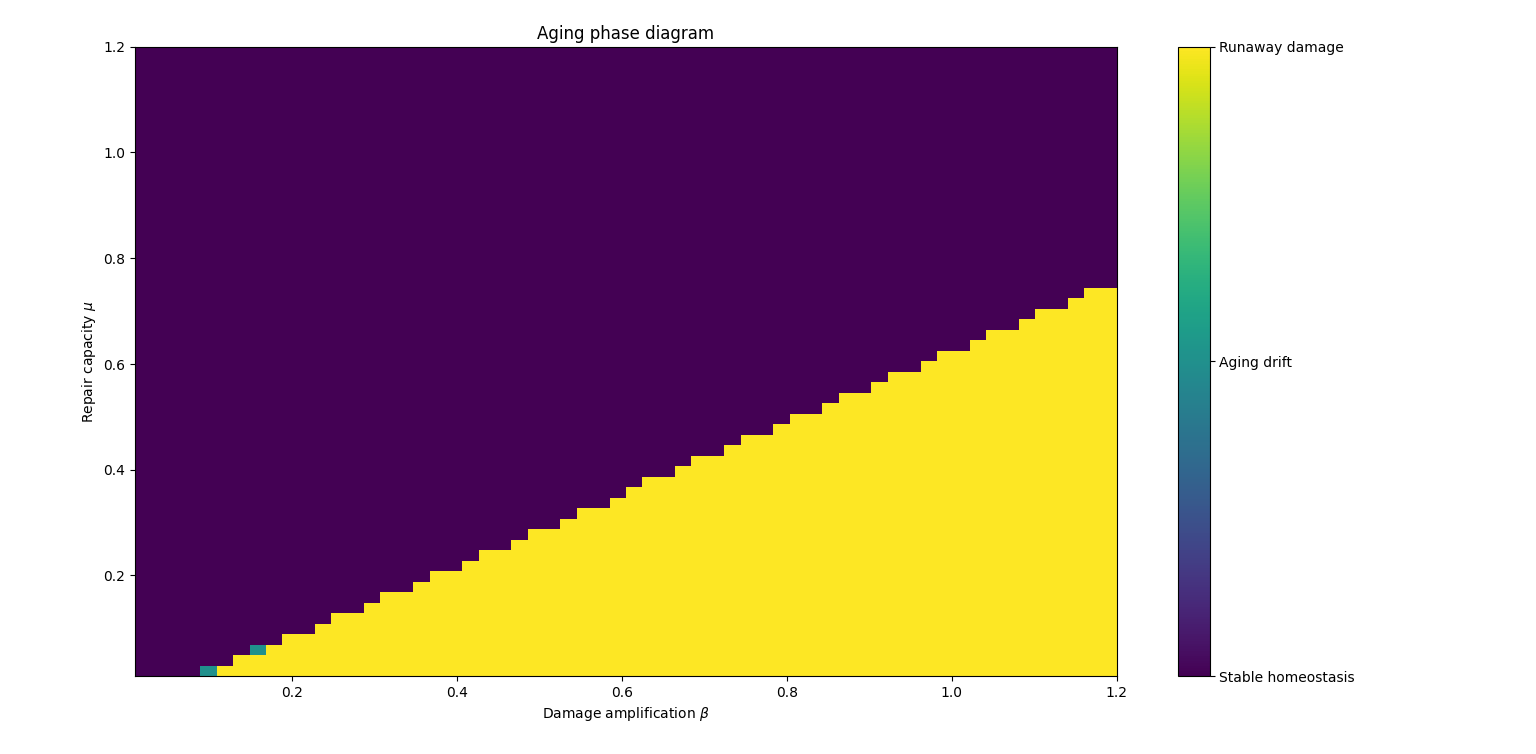}
\caption{\textbf{Phase diagram of aging dynamics.} Long-time behaviour of the damage variable $D(t)$ is classified across damage amplification $\beta$ (x-axis) and repair capacity $\mu$ (y-axis). For each grid point the initial condition $D(0)=D_0$ was integrated to $T_{\mathrm{final}}=200$ time units and classified as: stable homeostasis (tail variance $<10^{-3}$), aging drift (slow increasing tail slope), or runaway damage ($D(T_{\mathrm{final}})>10$). Colours denote regimes: stable homeostasis (purple), aging drift (teal), and runaway damage (yellow). The diagonal boundary indicates a critical trade-off between damage creation and repair: for each $\beta$ a minimum $\mu_c(\beta)$ is required to maintain bounded damage; conversely, for any $\mu$, there exists a maximum tolerated $\beta_c(\mu)$.}
\label{fig:one}
\end{figure}

To illustrate qualitative behaviour, a single-variable nonlinear damage model was used to produce a phase diagram over damage amplification $\beta$ and effective repair capacity $\mu$ (Fig.~\ref{fig:one}). At each grid point the initial condition $D(0)=D_0$ was integrated to $T_{\mathrm{final}}$ and classified as: (i) \emph{stable homeostasis} (tail variance below threshold), (ii) \emph{aging drift} (slow long-term increase without explosion across the horizon), or (iii) \emph{runaway damage} (fast divergence). The diagram shows a sharp boundary: for any fixed $\beta$ there exists a critical $\mu_c(\beta)$ such that $\mu>\mu_c(\beta)$ maintains boundedness, while $\mu<\mu_c(\beta)$ does not. Crucially, the boundary is diagonal: modest increases in $\mu$ cannot compensate for sufficiently large $\beta$, which demonstrates that increasing endogenous repair alone is not, in general, sufficient to arrest long-term damage growth unless information-limited lesion production is also addressed.

\begin{figure}
\centering
\includegraphics[width=0.95\linewidth]{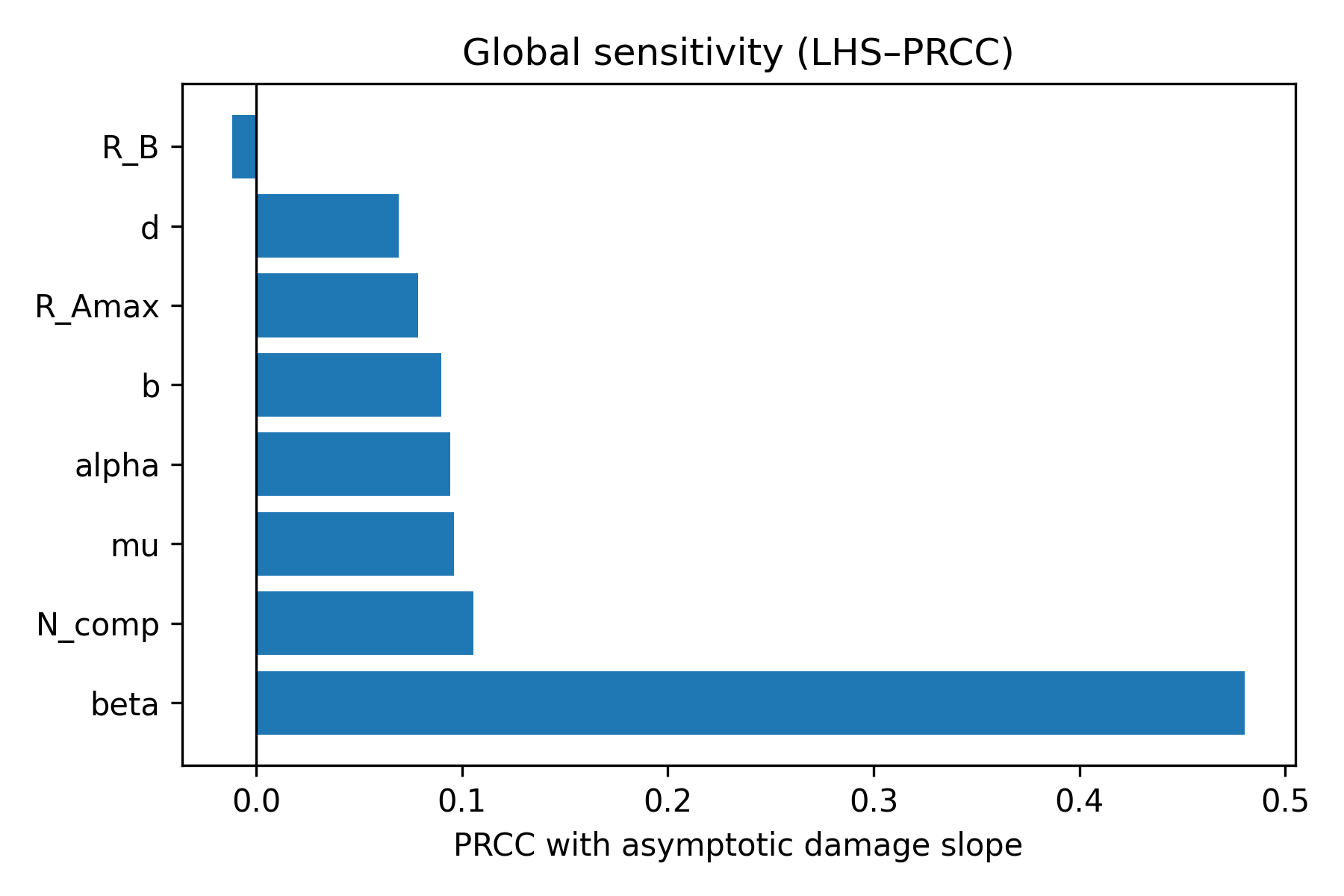}
\caption{\textbf{Global sensitivity of the asymptotic aging rate (LHS--PRCC).} Partial Rank Correlation Coefficients (PRCC) between model inputs and the long-term slope of total damage $D(t)$ from $N=3000$ Latin-hypercube samples. Deterministic trajectories were integrated to obtain asymptotic slopes; error bars show bootstrap 95\% confidence intervals obtained with 1000 resamples. PRCCs were FDR-corrected (Benjamini–Hochberg). The production rate of information-limited lesions ($\beta$) is the strongest positive predictor of the aging slope.}
\label{fig:two}
\end{figure}

A global sensitivity analysis was conducted to identify which parameters govern the asymptotic aging rate. Latin-hypercube sampling (N = 3000) spanned eight parameters. For each sample the deterministic ODEs produced the asymptotic slope of $D(t)$, and stochastic Gillespie ensembles quantified clonal risk proxies. Partial Rank Correlation Coefficients (PRCC) were computed on rank-transformed inputs and outputs controlling for other inputs. The PRCC vector for the asymptotic slope is shown in Fig.~\ref{fig:two} and numerically summarized as:
\[
\begin{aligned}
\mathrm{PRCC}(\alpha) &\approx +0.094, \\
\mathrm{PRCC}(\beta)  &\approx +0.480, \\
\mathrm{PRCC}(R_A^{\max}) &\approx +0.079, \\
\mathrm{PRCC}(R_B) &\approx -0.011, \\
\mathrm{PRCC}(\mu) &\approx +0.096, \\
\mathrm{PRCC}(b) &\approx +0.090, \\
\mathrm{PRCC}(d) &\approx +0.069, \\
\mathrm{PRCC}(N_\mathrm{comp}) &\approx +0.106.
\end{aligned}
\]

Two conclusions follow. First, the information-limited production rate $\beta$ is the dominant predictor of the long-term aging slope, exhibiting a moderate-to-strong positive PRCC. Second, the physiological repair ceiling $R_A^{\max}$ and stochastic clonal parameters have weak rank correlations with the asymptotic slope. The near-zero PRCC for engineered removal $R_B$ reflects threshold behaviour: unless $R_B$ crosses a critical value, small variations exert little incremental influence on the slope under uniform sampling (cf. Fig.~\ref{fig:one}). Bootstrap confidence intervals and FDR-corrected q-values validate the robustness of $\beta$'s dominance.

\section{Experimental observability of the effective damage removal rate}

A central requirement of any mechanistic aging theory is that its parameters correspond to experimentally measurable quantities. It was therefore examined whether the damage amplification parameter $\beta$ can be inferred from longitudinal observable data.

Consider a measurable functional quantity $S(t)$ governed by removal-limited dynamics
\[
S(t) = \exp(-\beta t),
\]
representing the persistence of function, survival probability, or biomarker signal under a constant effective removal rate. To mimic biological measurements, small observational noise was added to the trajectory. The parameter was then estimated directly from observable data using the logarithmic slope
\[
\hat{\beta} = -\frac{d}{dt}\log S(t).
\]

\begin{figure}
\centering
\includegraphics[width=0.95\linewidth]{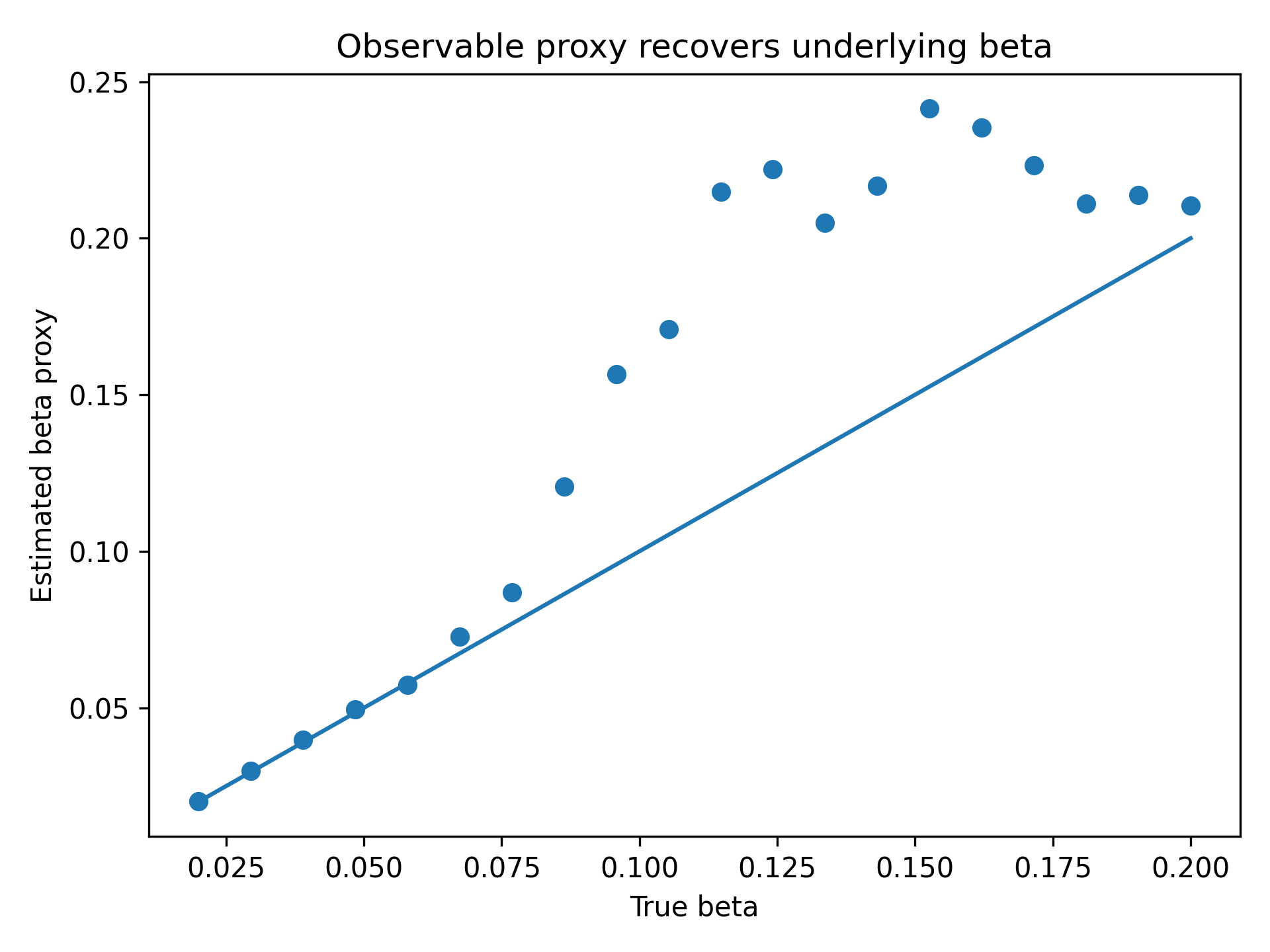}
\caption{\textbf{Identifiability of the effective damage removal rate.}
Synthetic longitudinal trajectories with measurement noise were generated under removal-limited dynamics $S(t)=\exp(-\beta t)$. The removal rate $\beta$ was inferred from the slope of $\log S(t)$. Estimated values closely matched the ground truth across the tested range, demonstrating that the model parameter corresponds to an observable biological quantity.}
\label{fig:three}
\end{figure}

Across a broad range of ground-truth values, the inferred estimates closely tracked the true parameter despite measurement noise (Fig.~\ref{fig:three}). This demonstrates that $\beta$ does not represent an abstract internal variable but corresponds to an experimentally identifiable effective damage-removal rate.

Consequently, the model generates directly falsifiable predictions: longitudinal functional decline trajectories should allow recovery of the underlying removal dynamics and quantitative comparison across interventions, genotypes, or species.

In practice, $S(t)$ may correspond to longitudinal functional capacity, frailty accumulation, clonal persistence, or other monotonic decline measures obtained in standard aging experiments.

\section{Operational implications and experimental tests}
The two-pathway control structure leads directly to operational predictions concerning intervention ordering, biomarker selection, and experimental design. If the systemic repair setpoint $S$ (hence proliferation $p(S)$) is raised prior to clonal curation and immune restoration, pre-existing clones can be amplified, producing transient increases in cancer risk \cite{Campisi2013,Jaiswal2017}. A simple cancer-risk proxy,
\[
C(t) \propto \frac{p(S)\,M(t)}{I(t)},
\]
with clone burden $M(t)$ and immune competence $I(t)$, quantifies this effect. Stochastic ensemble simulations demonstrate a higher fraction of runs exceeding a conservative clone-size threshold under a repair-first policy than under a de-risk-first policy (statistical comparison via Fisher exact test on exceedance counts). The recommended operational sequence is therefore: (i) surveillance and clonal de-risking, (ii) immune restoration, (iii) pulsed restoration of youthful systemic setpoints, and (iv) ongoing engineered removal to keep residual $B$-type lesions bounded.

\noindent\textbf{Operational prescription.} For safe implementation the framework prescribes the following sequence: (1) \emph{surveillance and de-risking} (increase observability and remove large clones) using longitudinal ctDNA VAFs, targeted single-cell mutation assays and imaging as needed; (2) \emph{immune restoration} guided by immune-repertoire sequencing and NK/T functional assays; (3) \emph{pulsed systemic repair activation} monitored by epigenetic clocks and autophagy/clearance flux assays; and (4) \emph{ongoing engineered maintenance} (repeat surveillance + targeted removal). These biomarkers provide immediate, operational estimates of information content $I(L;y)$, clone burden, and repair capacity required for adaptive control \cite{Horvath2013,Abbosh2017,Lu2020}.

The two-pathway decomposition classifies current approaches:

\textbf{Pathway A (increase $R_A$).} Exercise, caloric restriction and fasting mimetics, rapamycin and mTOR modulation, NAD$^+$ boosters, GLP-1 agonists, vagal neuromodulation, partial epigenetic reprogramming, and senolytics primarily increase endogenous repair or reduce $A$-type drivers \cite{Harrison2009,Longo2014,Ocampo2016,Justice2019}. These interventions improve transient and mid-term function but, per PRCC, are unlikely to arrest the asymptotic slope absent explicit control of $\beta$.

\textbf{Pathway B (reduce $\beta$ or increase $R_B$).} Targeted detection and removal of clones (ctDNA surveillance, targeted cytoreduction, engineered cellular therapies), mitochondrial replacement and allotopic strategies, in vivo gene correction (base/prime editing), cell replacement therapies, and matrix/AGE-removal approaches directly affect information-limited lesions and are, in principle, required to bound the long-term aging rate \cite{Li2018,Abbosh2017,Komor2016,Anzalone2019,Artika2020,Monnier2005}.

Programs that combine both pathways—surveillance and de-risking followed by repair restoration and sustained engineered maintenance—satisfy the sufficiency theorem in principle and are the natural translational targets.

Translational testing requires integrated biomarker panels. Suggested measures include multi-tissue epigenetic clocks and autophagy flux to estimate changes in $R_A$; longitudinal ctDNA VAFs and single-cell mutation assays to quantify $\beta$ and $\mu$; immune repertoire diversity and functional cytotoxicity assays to quantify $I(t)$. An in-silico trial using the calibrated stochastic model guides sample-size estimation and staging: if repair-first yields an estimated early-event probability $p_1$ and de-risk-first $p_2$, standard two-proportion sample-size formulas indicate required per-arm samples as a function of desired power and Type I error. Adaptive monitoring with early stopping for adverse signals (ctDNA increases, emergent driver VAFs) is recommended \cite{Barzilai2016,Horvath2013}.

The present formalism is intentionally abstract. Spatial heterogeneity, tissue-specific turnover, and ecological interactions among cell types can shift quantitative thresholds. The sufficiency theorem is conditional on the attainability of engineered removal at scale. Measurement error and imperfect biomarker observability affect identifiability; PRCC and profile-likelihood analyses partially mitigate these concerns and suggest focused panels for trial use. Ethical, regulatory and societal constraints—especially for gene editing and aggressive cytoreductive strategies—must be addressed before human trials \cite{Doudna2014,June2018}.

\paragraph*{Limitations}
The framework omits spatial heterogeneity, tissue-specific turnover differences, and multi-scale ecological interactions among cell populations. The sufficiency result assumes removal rates can be maintained over the relevant time horizon and does not address feasibility of specific therapeutic implementations. Quantitative thresholds should therefore be interpreted as structural rather than predictive.

\paragraph*{Conclusions}
Recasting aging as the simultaneous control of two logically distinct damage classes provides a compact, falsifiable engineering program: to maintain the boundedness of the modeled damage dynamics, it is necessary and sufficient to (i) sustain endogenous repair above production for regulatable lesions and (ii) bound or remove information-limited lesions. The phase diagram (Fig.~\ref{fig:one}) and global sensitivity analysis (Fig.~\ref{fig:two}) supply complementary evidence: a nontrivial phase boundary separates bounded from divergent regimes, and production of information-limited lesions predominates in determining the asymptotic aging rate. This framework clarifies why many interventions improve function without halting aging and offers an operational roadmap—detect, de-risk, restore, and engineer—that translates directly to trial design and safety monitoring \cite{Sanada2025,Levine2018}.

\section*{Data and Code Availability}
All simulations are fully reproducible from the public repository:
\url{https://github.com/Aging-supplementary-repository}.
The repository contains the exact parameter tables, random seeds, and scripts used to generate every figure.

\end{document}